\documentclass{article}
\usepackage{epsf,emulateapj}

\lefthead{IRWIN \& BREGMAN}
\righthead{SOFT X-RAY PROPERTIES OF LMXBs}
\slugcomment{Accepted by Astrophysical Journal Letters}

\begin{document}

\title{The Dependence of the Soft X-ray Properties of LMXBs on the Metallicity
of Their Environment}

\author{Jimmy A. Irwin and Joel N. Bregman}
\affil{Department of Astronomy, University of Michigan, \\
Ann Arbor, MI 48109-1090 \\
E-mail: jirwin@astro.lsa.umich.edu, jbregman@umich.edu}

\begin{abstract}
We determine the X-ray spectral properties of a sample of low-mass X-ray
binaries (LMXBs) which reside in globular clusters of M31, as well as five
LMXBs in Galactic globular clusters and in the Large Magellanic Cloud
using the {\it ROSAT} PSPC. We find a trend in the X-ray properties of the
LMXBs as a function of globular cluster metallicity. The spectra of LMXBs
become progressively softer as the metallicity of its environment increases.
The one M31 globular cluster LMXB in our sample which
has a metallicity greater than solar has spectral properties similar
to those of LMXBs in the bulge of M31, but markedly different from those
which reside in low metallicity globular clusters, both in M31 and the Galaxy.
The spectral properties of this high metallicity LMXB is also similar to
those of X-ray faint early-type galaxies. This lends support to the claim
that a majority of the X-ray emission from these X-ray faint early-type
galaxies results from LMXBs and not hot gas, as is the case in their
X-ray bright counterparts.
\end{abstract}


\section{INTRODUCTION} \label{sec:intro}

Nearly all that is known about the X-ray spectral properties of low-mass
X-ray binaries (LMXBs) has been derived at energies above 1 keV. The
simple reason for this is that most Galactic LMXBs lie in the Galactic
plane and are therefore heavily absorbed at soft X-ray energies by
Galactic hydrogen. Virtually nothing is known about the X-ray properties
of LMXBs below 0.5 keV. Two Galactic examples of LMXBs which lie
in directions of low hydrogen column densities (Hercules X-1 and MS1603+2600)
show strong X-ray emission at energies below 0.5 keV (Choi et al.\ 1997;
Hakala et al.\ 1998), in addition to the hard
(5--10 keV) emission normally attributed to LMXBs.

Fortunately, the bulge of M31 provides a relatively nearby (690 kpc)
laboratory for examining the soft X-ray properties of LMXBs. Since bulge
populations are known to be old, the X-ray sources found in the bulge
are not likely to be contaminated by high-mass X-ray binaries or
supernovae remnants. Supper et al.\ (1997) detected 22 X-ray sources
within $5^{\prime}$ of the center of the bulge with the {\it ROSAT} PSPC,
19 of which are most likely LMXBs with 0.1--2.0 keV X-ray luminosities
of $10^{36}-10^{38}$ erg s$^{-1}$ (the other three are supersoft sources).
Nearly all of these LMXBs exhibit strong very soft emission, much like
Hercules X-1 and MS1603+2600.

Furthermore, the integrated X-ray spectrum of the inner
$5^{\prime}$ of the bulge of M31, of which $\sim$80\% is resolved into
the 22 sources, strongly resembles the X-ray spectra of a class of early-type
galaxies that have very low X-ray--to--optical luminosity ratios
(Irwin \& Sarazin 1998a,b; hereafter IS98a,b).
Trends in the X-ray spectral properties of early-type glaxies were first
observed by {\it Einstein} by Kim, Fabbiano, \& Trinchieri (1992).
 Unlike X-ray bright early-type galaxies whose
X-ray emission is dominated by thermal emission from $\sim$0.8 keV gas,
these X-ray faint galaxies exhibit two-component (hard + very soft)
X-ray emission (Fabbiano, Kim, \& Trinchieri 1994; Pellegrini 1994;
Kim et al.\ 1996), with the hard component generally attributed to a
collection of LMXBs. The strong soft component was unsuccessfully attributed
to the integrated emission from M star coronae, RS CVn binary stars, and
supersoft sources (Pellegrini \& Fabbiano 1994; IS98b),
leaving only a warm (0.2 keV) ISM as a possible alternative.
However, LMXBs in the bulge of M31 demonstrated that LMXBs can also be a
significant source of very soft X-ray emission, and is a likely
explanation for the excess very soft X-ray emission in X-ray faint
early-type galaxies (IS98a,b) instead of a 0.2 keV ISM. The
X-ray--to--optical luminosity ratio of the bulge of M31 is comparable to
that of X-ray faint early-type galaxies, so LMXBs are also luminous
and/or numerous to account for the X-ray emission.

This seemingly simple solution is complicated by the fact that LMXBs exist
outside the disk of the Galaxy or the bulge of M31 which do not exhibit
strong very soft emission. Four of these LMXBs reside in Galactic globular
clusters, and another in the Large
Magellanic Cloud. This would seem to weaken the argument that
LMXBs are the source of the very soft emission in X-ray faint early-type
galaxies. However, these five examples differ from the rest in that they
reside in low metallicity environments. The metallicities of NGC~1851,
NGC~6624, NGC~6652, and NGC~7078 are 5\%, 43\%, 10\%, and 0.7\% solar,
respectively (Djorgovski 1993), and the iron abundance of the LMC is
$\sim$50\% solar (e.g., Hill, Andrievsky, \& Spite 1995).
It is possible that the metallicity of the environment in which an LMXB forms
has an effect on the X-ray properties of the binary.

To test this hypothesis we need a sample of LMXBs that reside in
environments that span a range of metallicities, and that also lie in
directions of reasonably low Galactic hydrogen column densities.
The globular cluster system of M31 is ideal for this. Metallicities
for most of M31's globular clusters have been determined optically, many
of which harbor LMXBs. In addition, the hydrogen column density towards M31
is not excessively high ($7 \times 10^{20}$ cm$^{-2}$). In this {\it Letter}
we present the spectral X-ray analysis of a sample of LMXBs which
reside in  M31 globular clusters using archival {\it ROSAT} PSPC data
to search for trends in the X-ray properties of LMXBs as a function of
the metallicity of their environment. We also determine the X-ray spectral
properties of the four Galactic globular cluster LMXBs with X-ray
luminosities $\ga 10^{36}$ erg s$^{-1}$ and LMC X-2.


\section{DATA REDUCTION} \label{sec:data}

Because of the large angular size of M31, the galaxy was imaged by the
{\it ROSAT} PSPC with many different pointings to encompass all the X-ray
emission. For our analysis, we only analyze the long ($>25$ kiloseconds)
observations. For each data set the error in the gain correction applied
by SASS as part of the conversion from detected pulse height to pulse
invariant channel (Snowden et al.\ 1995) was corrected using the FTOOL
pcpicor.

Each M31 X-ray source identified as being coincident with an M31 globular
cluster by Supper et al.\ (1997) was cross-referenced against 
Huchra, Brodie, \& Kent (1991) for an
estimate for the metallicity of the globular cluster.
If the source contained fewer than 250 X-ray counts
it was removed from the sample. Sources that fell outside the rib support
structure at $18^{\prime}$ were also removed. Table~\ref{tab:m31spec_fits}
lists the ID number of the X-ray source (taken from Supper et al.\ 1997),
an alternate ID taken from Huchra et al.\ (1991), and total X-ray
counts for each X-ray source in our sample.

For each LMXB, its spectrum was extracted from a circular aperture
centered on the source. The size of the extraction aperture varied with
the off-axis angle of the source to account for the point spread function
of the instrument. A locally-determined background was subtracted from the
spectra. The spectra were binned so that each channel contained at least
25 counts, and all channels below 0.1 keV and above 2.4 keV were ignored.
A similar procedure was used to extract spectra from LMXBs in NGC~1851,
NGC~6624, NGC~6652, NGC~7078 (M15), and LMC X-2.

\section{SPECTRAL FITTING} \label{sec:spectral}

\subsection{M31 Globular Cluster LMXBs} \label{subsec:m31_gc}

The spectra for the twelve M31 globular cluster LMXBs in our sample were fit
individually  with a variety of spectral models using XSPEC, and it was found
that absorbed thermal bremsstrahlung (TB), power law (PL), and blackbody models 
all fit the data equally well. We chose to let the photoelectric absorption
component (Morrison \& McCammon 1983) be a free parameter, since although the
Galactic hydrogen column density toward the direction of the LMXB is
known, it is not known how much M31 hydrogen the LMXB lies behind.
In most cases, when a blackbody model was used the best-fit column
density was well below (and statistically inconsistent with) the Galactic
value, so we discarded this model. The best-fit parameters for the TB and
PL models along with 90\% confidence levels
for one interesting parameter ($\Delta \chi^2 = 2.71$) are
shown in Table~\ref{tab:m31spec_fits} (all errors in this paper are quoted at
the 90\% level). Also shown are the minimum $\chi^2$ value,
the number of degrees of freedom, and the metallicity of the globular cluster
in which the X-ray source is embedded.

\subsection{Galactic and LMC LMXBs} \label{subsec:gal_gc}

Due to the relative proximity of the Galactic LMXBs plus LMC X-2, the
observations of these LMXBs yield a much greater number of X-ray counts
than the M31 globular cluster LMXBs. Consequently, simple one component
models are not adequate to describe the spectra of these objects as is
the case in the M31 LMXBs, which typically yield 1000 or fewer counts. This
illustrates the complexity of LMXB spectra at soft X-ray energies. In
order to compare fairly the properties between Galactic and M31 LMXBs,
we have analyzed only a small fraction of the PSPC data for the Galactic
LMXBs, such that each observation yields only $\sim$2000 counts. This number
was chosen to yield reasonable errors on the spectral parameters, while
still containing few enough counts to be directly comparable to the brighter
M31 LMXBs.

\begin{table*}[tbp]
\caption[M31 Spectral Fits]{}
\label{tab:m31spec_fits}
\begin{center}
\begin{tabular}{ccccccccccc}
\multicolumn{11}{c}{\sc Spectral Fits of M31 Globular Cluster LMXBs} \cr
\tableline \tableline
& & & &\multicolumn{3}{c}{TB Model} &&
\multicolumn{3}{c}{PL Model} \cr
\cline{5-7} \cline{9-11}
& Alternate & Total & & $N_H$ & $kT_{TB}$ & & & $N_H$ & & \\
Name & Name & Counts & $Z/Z_{\odot}$ & ($10^{20}$ cm$^{-2}$) & (keV) &
$\chi^2$/d.o.f. && ($10^{20}$ cm$^{-2}$) & $\Gamma$ & $\chi^2$/d.o.f. \\
\tableline
217 & 143-198 & 274 & 1.23 & 7.26$^{+15.06}_{-3.20}$ & 0.95$^{+1.64}_{-0.50}$ &
14.9/17 && 10.4$^{+34.1}_{-5.12}$ & 2.63$^{+2.29}_{-0.82}$ &
14.6/17 \\
228 & 153-000 & 820 & 0.83 & 7.81$^{+4.27}_{-1.95}$ & 1.70$^{+1.44}_{-0.60}$ &
30.0/28 && 9.47$^{+10.05}_{-2.73}$ & 2.01$^{+0.66}_{-0.34}$ &
31.7/28 \\
220 & 146-000 & 764 & 0.37 & 7.97$^{+4.08}_{-1.84}$ & 1.46$^{+0.99}_{-0.47}$ &
25.6/29 && 9.66$^{+9.56}_{-2.55}$ & 2.10$^{+0.64}_{-0.34}$ &
27.3/29 \\
73 & 005-52 & 2332 & 0.21 & 19.1$^{+6.33}_{-5.56}$ & 1.47$^{+1.13}_{-0.47}$ &
65.4/69 && 27.0$^{+10.9}_{-7.95}$ & 2.48$^{+0.62}_{-0.46}$ &
64.9/69 \\
282 & 225-280 & 730 & 0.20 & 6.21$^{+3.64}_{-1.74}$ & 3.57$^{+12.4}_{-1.84}$ &
19.4/25 && 7.08$^{+5.14}_{-2.33}$ & 1.64$^{+0.40}_{-0.33}$ &
19.8/25 \\
150 & 082-144 & 987 & 0.14 & 44.3$^{+18.2}_{-8.81}$ & 16.8$^{+\infty}_{-15.0}$ &
37.5/33 && 43.6$^{+24.9}_{-20.6}$ & 1.38$^{+0.99}_{-0.87}$ &
37.6/33 \\
122 & 045-108 & 1263 & 0.11 & 11.8$^{+8.87}_{-4.49}$ & 4.25$^{+76.9}_{-2.76}$ &
60.7/47 && 14.4$^{+14.6}_{-6.87}$ & 1.69$^{+0.82}_{-0.48}$ &
60.8/47 \\
247 & 185-235 & 435 & 0.09 & 5.59$^{+3.43}_{-1.53}$ & $> 4.15$ &
18.7/17 && 5.27$^{+6.09}_{-2.44}$ & 1.10$^{+0.50}_{-0.44}$ &
18.6/17 \\
349 & 386-322 & 1503 & 0.06 & 12.8$^{+8.03}_{-4.69}$ & 3.80$^{+14.2}_{-2.23}$ &
46.4/50 && 16.6$^{+12.7}_{-7.54}$ & 1.79$^{+0.69}_{-0.47}$ &
46.9/50 \\
318 & 375-307 & 5731 & 0.06 & 9.83$^{+1.60}_{-1.17}$ & 9.18$^{+18.0}_{-4.20}$ &
145.6/137 && 10.5$^{+2.25}_{-1.49}$ & 1.42$^{+0.16}_{-0.13}$ &
145.4/137 \\
205 & 135-193 & 1769 & 0.02 & 21.2$^{+7.82}_{-6.01}$ & 7.66$^{+\infty}_{-5.30}$ &
53.9/55 && 22.8$^{+11.1}_{-8.47}$ & 1.50$^{+0.55}_{-0.45}$ &
54.0/55 \\
158 & 086-148 & 485 & 0.02 & 5.50$^{+25.5}_{-1.55}$ & $> 0.81$ &
15.5/17 && 5.16$^{+42.1}_{-2.55}$ & 1.11$^{+2.39}_{-0.39}$ &
15.6/17 \\
\tableline
\end{tabular}
\end{center}
\end{table*}

As a check for consistency, five randomly-selected time intervals (each
of which yielded about 2000 counts) were analyzed from each observation
to search for any time dependence in the spectral properties. This was done
to ensure that the parameters of the models used to fit the spectra were
truly representative of time-averaged properties of the LMXB. Below, we
present the {\it median} best-fit parameters for the five time intervals
for each observation.

{\noindent \it NGC1851:} Based on a visual extinction of $A_V=0.06$ mag
(Djorgovski 1993), we have fixed the Galactic hydrogen column density
at $N_H=1.2 \times 10^{20}$ cm$^{-2}$ for both TB and PL models. We have
assumed $N_H=1.79 \times 10^{21} A_V$ cm$^{-2}$ (Predehl \& Schmitt 1995).
This produced a poor fit for the TB model and a marginal
fit for the PL model. However, Walker (1992) found the color excess to be
$E(B-V)=0.02 \pm 0.02$ for this cluster, so the absorption may be twice
the value quoted above given the error. Letting the absorption be free
led to acceptable fits with best-fit
parameters of $N_H=2.57^{+0.29}_{-0.25} \times 10^{20}$ cm$^{-2}$ and
$kT_{TB} > 19.4$ keV for the TB model, and
$N_H=1.90^{+0.49}_{-0.45} \times 10^{20}$ cm$^{-2}$ and $\Gamma=1.03 \pm{0.14}$
for the PL model. This is somewhat steeper than the value of $\Gamma=0.61$
found Verbunt et al.\ (1995) with {\it ROSAT} All Sky Survey data.

{\noindent \it NGC6624:} We have fixed the absorption
component at $N_H=1.56 \times 10^{21}$ cm$^{-2}$ ($A_V=0.87$ mag). Good
fits were found with a best-fit parameter of $kT_{TB}=3.77^{+2.72}_{-1.23}$
for the TB model and $\Gamma=1.57 \pm 0.14$ for the PL model.
Again, this is slightly steeper than the value of $\Gamma=1.36$ of
Verbunt et al.\ (1995).

{\noindent \it NGC6652:}
The absorption was fixed at $N_H=5.55 \times 10^{20}$ cm$^{-2}$
($A_V=0.31$ mag). A marginal fit ($\chi_{\nu} \sim 1.7$) was found for a
TB model with an unreasonably high lower limit on the temperature of
$kT_{TB} > 100$ keV. However, a good fit was obtained with a PL model
with $\Gamma=0.80 \pm 0.10$. Verbunt et al.\ (1995) found a value of
$\Gamma=0.72$.

{\noindent \it NGC7078:}
The absorption was fixed at $N_H=1.97 \times 10^{20}$ cm$^{-2}$
($A_V=0.11$ mag). A TB model gave a very poor fit to the data. A better fit
was obtained with a PL model with $\Gamma=0.12 \pm 0.10$.

{\noindent \it LMC X-2:} The column density was a free parameter.
The best-fit values were $N_H=1.30^{+0.62}_{-0.44} \times 10^{21}$ cm$^{-2}$
and $kT_{TB}=2.47^{+3.91}_{-1.08}$ for the TB model, and
$N_H=1.87^{+0.62}_{-0.44} \times 10^{21}$ cm$^{-2}$ and
$\Gamma=2.06^{+0.59}_{-0.53}$ for the PL model. Estimates made from color
excess maps of the LMC by Schwering \& Israel (1991) give a column density of
$\sim$$10^{21}$ cm$^{-2}$, in rough agreement with the X-ray measurements.

\section{DISCUSSION} \label{sec:discussion}

From Table~\ref{tab:m31spec_fits} the correlation between the best-fit
TB temperature or PL exponent and the metallicity is evident.
Source 217, which has the highest metallicity of the LMXBs in the sample
also has the lowest measured temperature ($kT=0.95$ keV)
or steepest power law exponent ($\Gamma=2.63$).
Conversely, the lower metallicity LMXBs have higher temperatures
or flatter power law exponents.
In some cases, the best-fit $N_H$ value is considerably higher than the
Galactic value, indicating that the LMXB spectrum is being absorbed by
hydrogen in the disk of M31. In other cases the derived $N_H$ value is
consistent with the Galactic value in that direction. This trend with
metallicity is supported by the four Galactic globular cluster LMXBs and
LMC X-2. The two LMXBs in higher metallicity environments (NGC~6624
and LMC X-2) have softer spectra than the three low metallicity LMXBs.

Previous studies (IS98a,b) have concluded that the
X-ray properties of a subclass of early-type galaxies with very low
X-ray--to--optical luminosity ratios are similar to those of LMXBs
in the Galaxy and the bulges of M31 and NGC~1291. In these studies, two
X-ray ``colors" (ratio of counts in three X-ray bands) were used to
characterize the X-ray emission. For a comparison to those studies, we
compute the same colors from the best fit TB and PL fits described
above. The two X-ray colors, C21 and C32, are defined as
\begin{equation} \label{eq:c21}
{\rm C21} =
\frac{\rm counts~in~0.52-0.90~keV~band}{\rm counts~in~0.11-0.41~keV~band}
\, ,
\end{equation}
and
\begin{equation} \label{eq:c32}
{\rm C32} =
\frac{\rm counts~in~0.91-2.02~keV~band}{\rm counts~in~0.52-0.90~keV~band}
\, .
\end{equation}

The absorption-corrected colors derived from the best-fit models for
the twelve LMXBs are shown in Figure~\ref{fig:brem} and
Figure~\ref{fig:power} for the TB and PL models,
respectively. The errors shown are the colors derived from spectral models
using the 90\% upper and lower limits on the temperature or power law
exponent. The M31 globular cluster sample has been
broken into three groups based on metallicity: $Z \le 0.2$ $Z_{\odot}$,
0.2 $Z_{\odot} < Z <$ $Z_{\odot}$, and $Z >$ $Z_{\odot}$.
The colors of the four Galactic LMXBs and LMC X-2 are also shown, although
NGC~6652 and NGC~7078 have been omitted in the TB case since this model
did not produce an acceptable fit to the spectra.
In the TB case, the temperature of some of the low metallicity LMXBs
was unconstrained, so the colors given are those predicted by a
bremsstrahlung temperature of 200 keV (the highest temperature allowed
by XSPEC). Also shown are the colors for the bulge of M31
and many X-ray faint early-type galaxies
(taken from IS98b). The error bars on these quantities
have been omitted for clarity.

The segregation of the colors with metallicity of the globular cluster is
clear, although the errors are rather large in the power law case. As the
metallicity of the cluster decreases the colors increase, indicating a
hardening of the spectra. The lone LMXB in the sample that resides in
a cluster with a metallicity greater than solar has colors very similar to
that of the bulge of M31, which has a metallicity of about twice solar
(Bica, Alloin, \& Schmidt 1990). As mentioned above, the X-ray emission from
the bulge of M31 is dominated by LMXBs. The three moderate metallicity M31
LMXBs as well as NGC~6624 and LMC X-2 occupy a region of C21-C32 space above
and to the right of the high metallicity LMXB and the bulge of M31.
The lowest metallicity LMXBs lie further still to the right and above.
The three low metallicity Galactic globular cluster LMXBs have especially
hard colors in the PL case.

For both the TB and PL case, the colors of the high metallicity M31 LMXB
(Source 217) is consistent with those of the X-ray faint early-type galaxies.
As mentioned above, the LMXBs in the bulge of M31 give X-ray--to--optical
luminosity ratios comparable to those of the X-ray faint galaxies. The colors
of the high metallicity M31

\centerline{\null}
\vskip2.65truein
\includegraphics{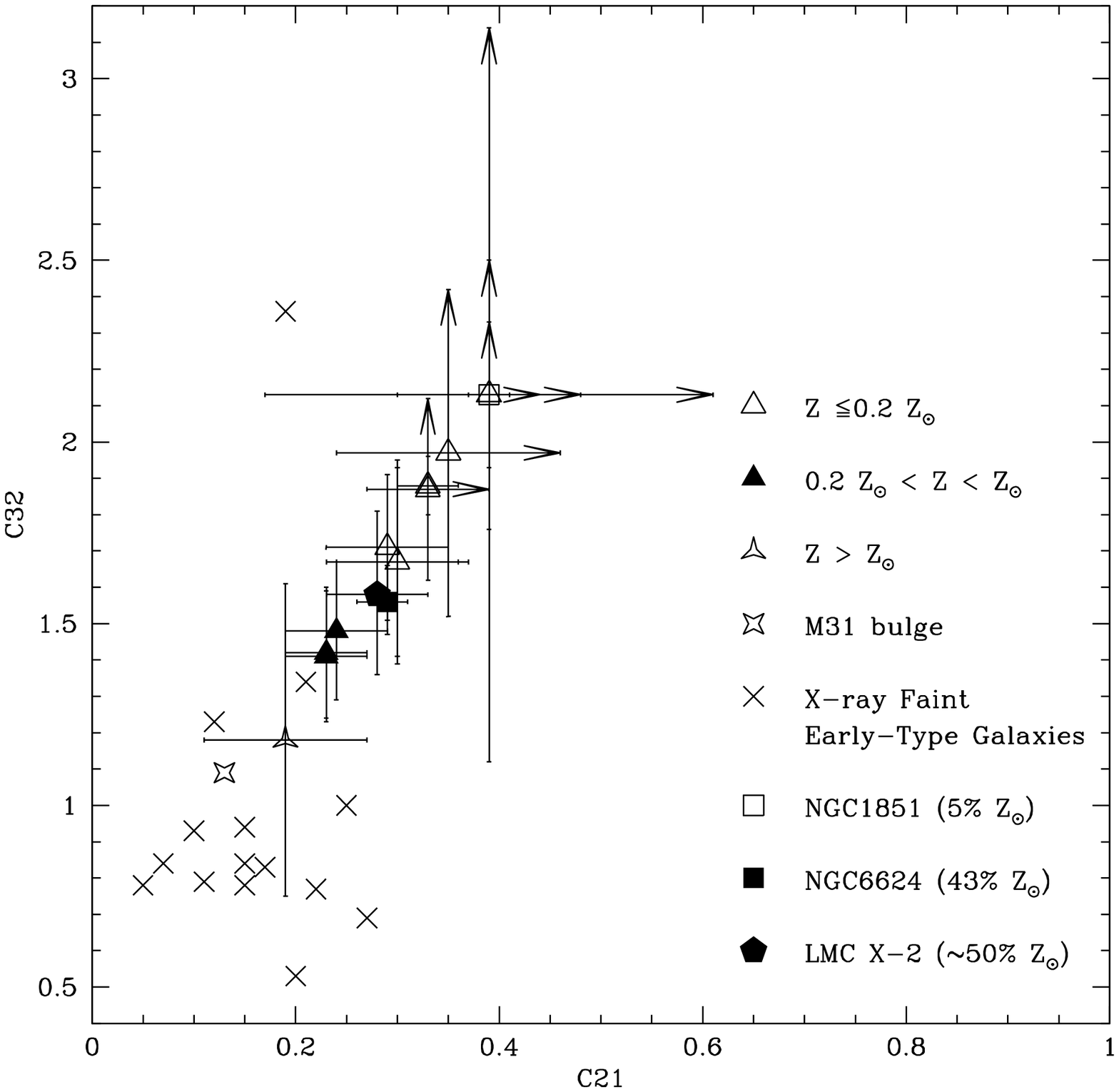}
\figcaption{
C21 vs.\ C32 plot of the best-fit TB models of the twelve M31 globular cluster
LMXBs, NGC~1851, NGC~6624, and LMC X-2. Also shown are the colors of
the bulge of M31 and many X-ray faint early-type galaxies
(taken from IS98a). Arrow indicate that only lower limits
were found for the temperatures. The trend of X-ray colors with the
metallicity is evident. Note that the high metallicity M31 LMXB has colors
similar to those of the bulge of M31 and the X-ray faint galaxies.
\label{fig:brem}}

\vskip0.1truein

\centerline{\null}
\vskip2.65truein
\includegraphics{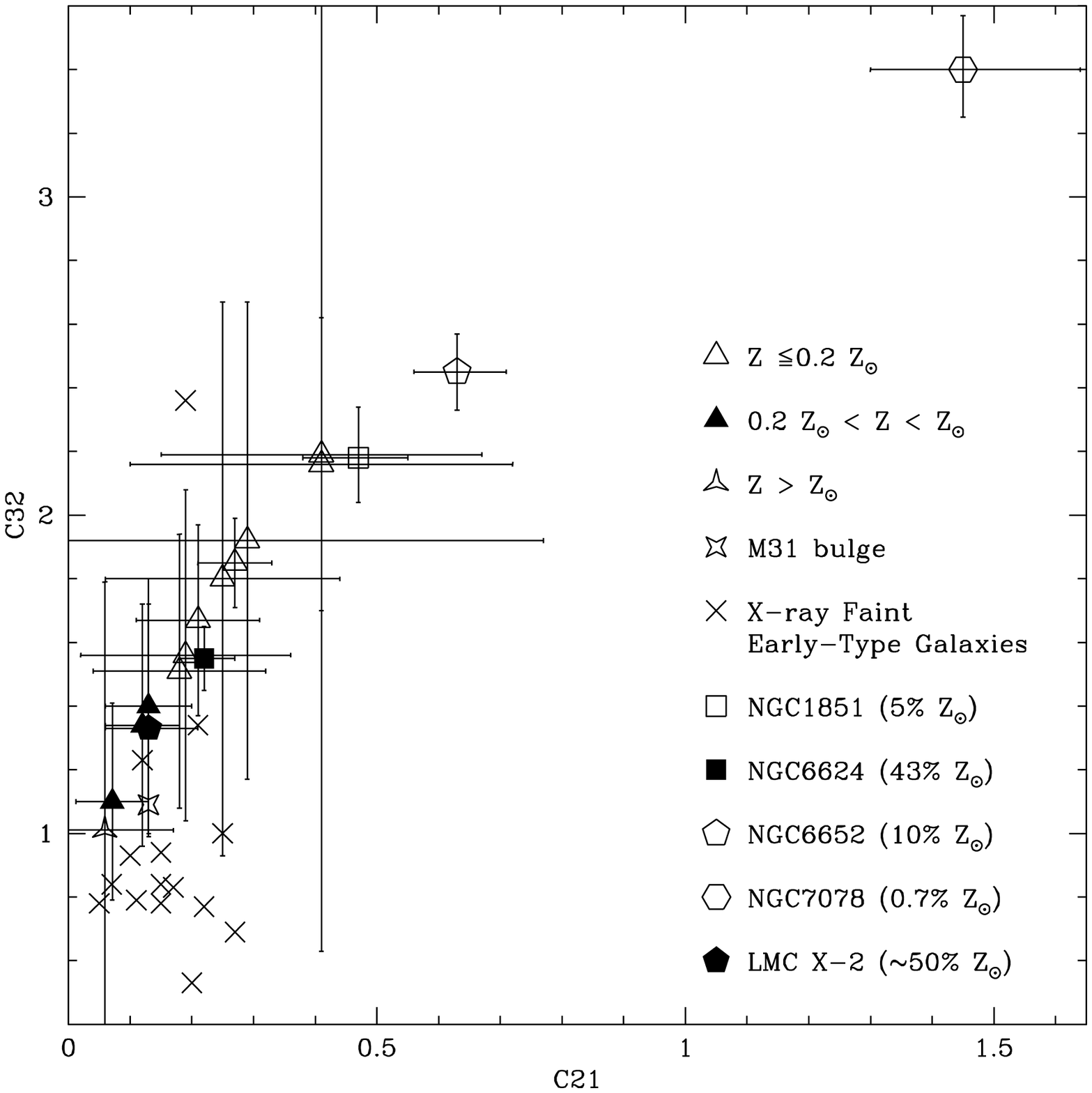}
\figcaption{
Same notation as in Figure~\protect\ref{fig:brem}, but this time including
the colors of NGC~6652 and NGC~7078 for the PL case. Although the errors
are larger than in the TB case, the trend of X-ray colors with the
metallicity is still evident.
\label{fig:power}}

\vskip0.1truein

\noindent globular cluster LMXB adds further
evidence that LMXBs can produce the spectral characteristics observed in X-ray
faint early-type galaxies. Given the high central metallicities of early-type
galaxies, the LMXBs in these systems should more closely resemble LMXBs
in high metallicity systems such as the bulge of M31 or Source 217, and
not LMXBs in low metallicity systems such as those in NGC~1851, NGC~6652,
and NGC~7078.

Previous studies (i.e., Davis \& White 1996) have found a correlation
between ISM temperature and metal abundance, by assuming that the
X-ray spectra of the X-ray faintest galaxies could be described by a
single component, zero metallicity Raymond-Smith model with
$kT\sim 0.6$ keV. Although this model is not excluded by the {\it ROSAT}
PSPC data, a subsequent {\it ASCA} study of at least one X-ray faint galaxy
excluded this model (Kim et al.\ 1996). If a majority of the X-ray
emission from the X-ray faintest galaxies is stellar in nature, this calls
into question the relation between ISM temperature and abundance, as
well as the relation between ISM temperature and stellar velocity
dispersions.

Finally we note that the small discrepancy in C32 between the bulge of
M31 and the mean of the X-ray faint early-type galaxies may be caused
by the presence of small amounts of warm interstellar gas in the X-ray
faint systems, although the gas is not the dominant X-ray emission
mechanism as is the case in X-ray bright galaxies. There is
evidence that the amount of ISM increases with increasing
$L_X/L_B$. As a comparison, X-ray
bright early-type galaxies whose emission is dominated by $\sim$0.8 keV
gas have X-ray colors in the range (C21,C32$)=(0.5-1, 0.8-2)$, well
separated from the X-ray faint early-type galaxies (IS98b).

\section{A FEW WORDS OF WARNING} \label{sec:warning}

It should be stressed that the low energy spectra of LMXBs are quite
complicated and cannot in general be described by a single TB or PL
model. The good fits obtained here are solely the result of having a
paucity of X-ray counts for the M31 sample. For the well-observed Galactic
LMXB sample, even two component fits did not always produce a good fit.
However, there is no reason to believe that the simple models used here
should lead to misleading results regarding the trend of the X-ray colors
with metallicity.

Another issue not addressed here is the the presence of absorbing material
intrinsic to the LMXB. For the five Galactic/LMC LMXBs as well as eight of
the twelve M31 LMXBs, the best-fit absorption value is consistent
with what is expected from intervening material in our own Galaxy.
We have assumed that those showing excess absorption do so because they
lie behind absorbing material in the disk of M31. Although it is possible
that the excess absorption is from material intrinsic to the LMXB, the removal
of these four LMXBs will not alter the conclusions drawn here. Three of the
four LMXBs have a low metallicity and already have hard colors, so using a
lower value for the line-of-sight (non-intrinsic) absorption will only
make their colors harder. Furthermore, inspection of the raw C32 color
{\it uncorrected} for absorption shows the same trend with metallicity.
The trend in C21 is lost, though, since this color is highly dependent
on absorption.


\acknowledgments

This research has made use of data obtained through the High Energy
Astrophysics Science Archive Research Center Online Service,
provided by the NASA/Goddard Space Flight Center.
This work has been supported by NASA grant NAG5-3247.

\end{document}